\documentstyle[epsfig]{l-aa}

\begin{document}

\def\ltsima{$\; \buildrel < \over \sim \;$}
\def\simlt{\lower.5ex\hbox{\ltsima}}

   \thesaurus{ 4(11.09.1: IC 4329A;  
                11.19.1; 
                13.25.2)} 
   \title{X-ray spectral components from a broad band BeppoSAX observation 
of the Seyfert galaxy IC 4329A}

   \author{G. C. Perola
          \inst{1}
    \and G. Matt 
          \inst{1}
\and M. Cappi
          \inst{2}
\and D. Dal Fiume
          \inst{2}
\and F. Fiore
          \inst{3}
\and M. Guainazzi
          \inst{4}
\and T. Mineo
          \inst{5}
\and S. Molendi
          \inst{6}
\and F. Nicastro
          \inst{7}
\and L. Piro
          \inst{8}
\and G. Stirpe
          \inst{9}
          }

   \offprints{G.C. Perola, perola@fis.uniroma3.it}

   \institute{Dipartimento di Fisica, Universit\`a degli Studi ``Roma Tre", 
              Via della Vasca Navale 84, I--00146 Roma, Italy
   \and Istituto Tecnologie e Studio Radiazioni Extraterrestri, CNR, Via
              Gobetti 101, I--40129 Bologna, Italy
      \and Osservatorio Astronomico di Roma, Via dell'Osservatorio,
        I--00044 Monteporzio Catone, Italy
   \and Astrophysics Division, Space Science Department of
        ESA, ESTEC, Postbus 299, NL-2200 AG Noordwijk, The Netherlands
   \and Istituto di Fisica Cosmica ed Applicazioni all'Informatica, C.N.R., 
        Via U. La Malfa 153, I--90146 Palermo, Italy
   \and Istituto di Fisica Cosmica ``G. Occhialini", C.N.R., Via Bassini 15,
        I--20133 Milano, Italy
   \and Harvard-Smithsonian Center for Astrophysics,
		60 Garden st., Cambridge MA 02138 USA 
   \and Istituto di Astrofisica Spaziale, C.N.R., Via Fosso del Cavaliere,
                I--00133 Roma, Italy
   \and Osservatorio Astronomico di Bologna, Viale Berti Pichat 6/2, 
  	I--40127 Bologna, Italy
   }

   \date{Received / Accepted }

   \maketitle

\markboth{G.C. Perola et al.: BeppoSAX observation of IC 4329A}{}

   \begin{abstract}
From the spectral analysis of a broadband (0.1--200 keV) BeppoSAX observation
of the Seyfert 1 galaxy IC 4329A, the main results obtained are:
a) the amount of reflection, together with the intensity of the iron K line,
indicate a geometry with a solid angle substantially less than 2$\pi$;
b) the power law is affected by a cut off with $e$-folding energy about
270 keV, the fourth individual object so far where this property
has been firmly detected; c) two absorption features at about 0.7 and 1 keV
are found, the first corresponding to a blend of
O~{\sc vi} and O~{\sc vii},
the other to a combination of FeL and NeK edges. Compared to an
earlier ASCA observation, when the source was 30\% fainter,
the values of both the relative
amount of reflection and the warm absorber ionization degree
are significantly lower: the comparison is suggestive of sizeable
delay effects in this object, due to geometrical factors in the reflection,
and to relaxation to equilibrium states in the ionization of the absorber.
   \end{abstract}

%

\section{Introduction}

Starting from EXOSAT and Ginga observations, it has become progressively
clear that the X--ray spectra of Seyfert 1 galaxies are rather complex
(see, e.g., Mushotzky et al. 1993, and ref. therein), 
with the effects of radiative 
transfer in the environment of the central engine complicating the
view of the primary emission. The latter is empirically
described as a power law, which, according to OSSE/CGRO observations 
(Gondek et al. 1996, Zdziarski et al. 1995), appears to
break at high energies in a manner that can be described to a first 
approximation by an exponential cutoff, although the value of the $e$-folding
energy in individual objects remains to be properly constrained.
At energies below about 1--2 keV the incidence
of a further continuum component in excess to the power law, the so--called
soft X--ray excess, is not yet completely well assessed,
with ASCA data showing it to be rather less common
(e.g. Reynolds 1997) than in ROSAT data (e.g. Walter \& Fink
1993). In fact this issue had a significant turn after
the discovery with ROSAT and, more accurately, with ASCA of the rather
common incidence of a warm absorber: in essence, both types of features
are determined with less ambiguity the better the slope of the power
law is constrained, a task which ASCA, in addition to the higher
energy resolution, could achieve more directly than ROSAT 
because of its energy coverage 
up to 10 keV. To complicate the matter there is another,
seemingly ubiquitous,  continuum component, which affects
the power law slope determination in the ASCA band (as it previously 
did in the HEAO1 and EXOSAT bands) and peaks beyond that band, around
30 keV. This component, discovered with Ginga, is interpreted as
Compton reflection of the power law photons off thick matter that can be
the accretion disk itself and/or the torus envisaged in the unified
Seyfert 1 and 2 scenario. This component is typically accompanied
by the iron K fluorescence line, discovered in fact before the
reflection. The accuracy achieved on the strength
of the reflection is then relevant to the determination of the power law
slope and, consequently, of the soft excess and/or
the warm absorber parameters. 
Observationally the situation is further complicated by the well known
variability of these objects, such that combining data from satellites
covering different energy ranges and collected at different times 
leads to further ambiguities. 

The Narrow Field Instruments aboard BeppoSAX offer
for the first time the opportunity to cover simultaneously,
and with a reasonably good spectral
resolution, the band from 0.1 to 200 keV, hence to determine the 
various spectral parameters, especially the broad ones, with
better confidence than could be done before. 
This paper reports results obtained as part of a broad band spectral survey
of Seyfert 1s, with 2--10 keV fluxes greater than
about 1--2 mCrab. After the observation of NGC 4593
described in Guainazzi et al. (1999a) here we present
the observation of the brightest source in the sample,
IC 4329A. In Sect. 2 the earlier X--ray studies
are summarized. In Sects. 3 and 4, the data reduction and the spectral 
analysis are respectively presented. In Sect. 5 these
results are commented on and compared with previous ones, and
the conclusions are drawn in Sect. 6.

\section{IC 4329A in X--rays}

IC 4329A is classified as a Seyfert 1.0 galaxy (Whittle 1992) at
z=0.016 (Wilson \& Penston 1979). The 2--10 keV flux records 
in the literature show a relatively
modest variability on timescales of hours--days, and fall within
50\% of 1.0$\times$10$^{-10}$ erg cm$^{-2}$ s$^{-1}$ 
(corrected for absorption), 
corresponding to L(2--10 keV) = 10$^{44}$ erg s$^{-1}$ ($H_0$=50 km 
s$^{-1}$ Mpc$^{-1}$). However, from inspection of the 2--10 keV light
curve from the ASM instrument onboard Rossi XTE (quick look results
provided by the ASM/RXTE team) it appears that this object undergoes
occasional outbursts by up to a factor of about 10, with timescales
of the order of a few days. Furthermore a factor 3 decline in about four
days in visible in the light curve from the January 1993 OSSE/CGRO
observation, albeit in the quite different band from 50 to 150 keV
(Fabian et al. 1993).

The spectrum displays neutral absorption at low energies, due to a gas
column about ten times greater than the galactic value 
N$_{Hg}$=4.5$\times$10$^{20}$
cm$^{-2}$ (Elvis et al. 1989), conceivably associated with the 
disk of the host galaxy, which is oriented nearly edge--on (Petre et al.
1984). 

This object is one of the first AGN where a reflection component
was individually detected with Ginga (July 8--10 1989 observation;
Piro et al. 1990, Nandra \& Pounds 1994): the amplitude of this
component relative to the power law  was found to be roughly consistent with
an $\Omega$=2$\pi$ slab of reflecting matter, and the intensity 
of the iron K line simultaneously measured had EW$\simeq$110 eV 
from a gaussian fit. A time resolved spectral analysis of the
same data by Fiore et al. (1992) presents tentative evidence of a time 
lag by more than three days in the response
of the reflection component to an increase of the 
power law by 20\% in three days. Contemporaneous observations with
ROSAT and OSSE/CGRO in January 1993 (Madejski et al. 1995)
gave for the reflection component results, with large errors, consistent
with those obtained by Ginga, providing in addition a lower limit of
about 100 keV for the $e$--folding energy of the power law. The August 15
1993 ASCA observation has been analyzed in several papers
(Mushotzky et al. 1995, Cappi et al. 1996, Nandra et al. 1997, 
Reynolds 1997, George et al. 1998). The detailed analysis by Cappi et al. 
(1996), for what concerns the reflection,
provides results surprisingly at variance with those obtained 
with Ginga: despite the flux level being only 30\% lower, 
the best fit relative normalization of the reflection component 
is about three times stronger, while the gaussian fit of the
iron line yields about the same EW (89$\pm$34 eV at 90\% confidence
for one interesting parameter, i. p.). This result is hard to reconcile 
with the idea that both the line and 
the reflection arise from the same matter, and we shall return to
it in Sect. 5. 
 
At low energies Madejski et al. (1995) discovered with ROSAT
the existence either of the ionized edge imprinting by an
additional warm absorber, or of a soft excess. The ambiguity
was convincingly resolved by Cappi et al. (1996), who found
in the ASCA data evidence of two rather strong edges, one
consistent with O~{\sc vii}, the other with O~{\sc viii} (this aspect is
investigated also in Reynolds 1997 and George et al. 1998, see Sect. 5).

\section{Observations, data reduction and temporal analysis}

A general description of the Italian--Dutch satellite BeppoSAX
can be found in Boella et al. (1997a). The observations were
made with the four, coaligned Narrow Field Instruments:
the two imaging instruments covering, respectively, the
band 0.1--10 keV (LECS, Low Energy Concentrator Spectrometer,
Parmar et al. 1997) and the band 1.8--10.5 keV (MECS, Medium
Energy Concentrator Spectrometer, Boella et al. 1997b); the two collimated
instruments, covering respectively the band 4--120 keV (HPGSPC,
High Pressure Gas Scintillation Proportional Counter, Manzo et
al. 1997) and the band 13--200 keV (PDS, Phoswich Detector System,
Frontera et al. 1997), with their collimators operating in the rocking
mode (time--on = time--off = 96 s) to monitor the background throughout
the observation. The data from the HPGSPC, an instrument tuned
for spectroscopy of very bright sources, provided
in our case barely significant constraints 
and will therefore not be considered in this paper. Due to rather subtle
problems that remain in the exploitation for spectral analysis
of LECS data above 4 keV, we shall use only data in the range 0.1--4.0 keV.

In Table 1 the start date of the observation, the net exposure times
and the net average count rates are given. The exposure time in the
LECS is much shorter than in the MECS because the former instrument
is operated only during the night--time fraction of each orbit.
The reduction procedures and screening criteria used to produce
the linearized and (between the two MECS units) equalized event files
are standard and have been described in Guainazzi et al. (1999a).
In particular, of the two options available for the PDS, we used
the Rise Time selection criterion. 

\begin{table*}
\centering
\caption{Observation epoch and mean count rates. The count rate for the MECS 
refers to two units}
\vspace{0.05in}
\begin{tabular}{|c|cc|cc|cc|}
\hline
~ & ~ & ~ & ~ & ~ & ~ & ~\cr
Start date  & LECS  & LECS & MECS & MECS  & PDS & PDS\cr
\hline
~ & t$_{\rm exp}$ & CR~(0.1-10~keV) & t$_{\rm exp}$ & CR~(1.8-10.5~keV) 
& t$_{\rm exp}$ & CR~(13-200~keV) \cr
~ & s & cts~s$^{-1}$ & s & cts~s$^{-1}$ & s & cts~s$^{-1}$ \cr
\noalign {\hrule}
\hline
~ & ~ & ~ & ~ & ~ & ~ & ~\cr
1998-Jan-02~(9h~1m~16s~UT) & 25097 & 0.696$\pm$0.005 & 81826 
& 1.551$\pm$0.004 & 75214 & 
0.983$\pm$0.015 \cr
~ & ~ & ~ & ~ & ~ & ~ & ~\cr
\hline
\end{tabular}
\end{table*}

The spectral counts in the imaging instruments were extracted from
circular regions of radius 4 arcmin (MECS) and 8 arcmin (LECS)
around the source centroid, and the background subtraction was 
performed using spectra from blank sky event files in the same
position of the detectors. The background count rate
used is 2.74$\times$10$^{-2}$ c/s in the LECS, 
6.95$\times$10$^{-3}$ c/s (2 units) in the MECS, and the error
on the net counts (in Table 1) is dominated by the source statistics. 
Spectra and light curves from the PDS were obtained from direct subtraction
of the off-- from the on--source products, and the error on the net
counts (in Table 1) is dominated by the background statistics.

In the field of view of the PDS, 1.3$^{\circ}$ FWHM, care must be taken
of sources that could contaminate the target signal. 
IC 4329A belongs to the cluster
of galaxies A 3574, which is a comparatively very weak X--ray emitter
(Pierre et al. 1994). We integrated its surface brightness in the
ROSAT pointed observation described in Madejski et al. (1995)
to obtain a rough estimate of the flux: in the 2--10 keV range
(with an hypothetically large value of the temperature, $kT$=7 keV)
it is about 30 times lower than that of IC 4329A, hence its 
contribution in the PDS is totally negligible. The same conclusion
holds for the point source named S3 by Madejski et al. (1995),
which lies at about 12 arcmin from the target, and that we find
in the MECS image with a 2--10 keV flux of about
8.4$\times$10$^{-13}$ erg cm$^{-2}$ s$^{-1}$, 
more than 100 times fainter than IC 4329A.
Another source, associated with the elliptical galaxy IC 4329 
located only 3 arcmin away from
IC 4329A, might contaminate the signal of our target extracted from the
LECS. Madejski et al. (1995) attribute to this source a thermal 
plasma spectrum with $kT$=0.87 keV, and a flux 
6.5$\times$10$^{-13}$ erg cm$^{-2}$ s$^{-1}$ in the 0.1--2 keV range: 
from Fig. 6 of their paper we expect
the contamination by this source to be possibly relevant only
below 0.4 keV (but see Sect. 4). 

The detailed study of the observed time variability 
goes beyond the scope of this paper, but some information is presented 
to justify the spectral analysis of the integrated spectra. 
Fig. 1 illustrates light and hardness ratio (HR) curves.
The object underwent
important intensity variations, but the 
$\chi^2$ test applied to the strings of HR values (Table 2)
shows that the evidence of deviations from a
constant value is marginal overall: we conclude that 
spectral variations are unlikely to introduce, 
in the analysis on the integrated counts,
an important bias in the estimate of the parameters. We shall,
though, take care 
of the inhomogeneity in the time coverage between
LECS and MECS, by letting the relative normalization of the
two instruments as a free parameter. The normalization of the 
PDS to the MECS will instead be held fixed.

\begin{table*}
\centering
\caption{Hardness ratios: mean values and $\chi^2$ for a constant value}
\vspace{0.05in}
\begin{tabular}{|cc|cc|cc|}
\hline
~ & ~ & ~ & ~ & ~ & ~\cr
HR(${1.8-4 \over 0.1-1.8}$)$^a$ & $\chi^2_{\nu}$/d.o.f. &
HR(${4-10.5 \over 1.8-4}$)$^b$ & $\chi^2_{\nu}$/d.o.f. &
HR(${13-100 \over 1.8-10.5}$) & $\chi^2_{\nu}$/d.o.f.  \cr
~ & ~ & ~ & ~ & ~ & ~\cr
\hline
\noalign {\hrule}
\hline
~ & ~ & ~ & ~ & ~ & ~\cr
1.04 & 1.27/28 & 0.74 & 0.83/28 & 0.61 & 1.31/28 \cr
~ & ~ & ~ & ~ & ~ & ~\cr
\hline
\end{tabular}
\begin{tabular}{c}
$^a$based on LECS data only\par
~~~~~~$^b$based on MECS data only
\end{tabular}
\end{table*}

\begin{figure}
\epsfig{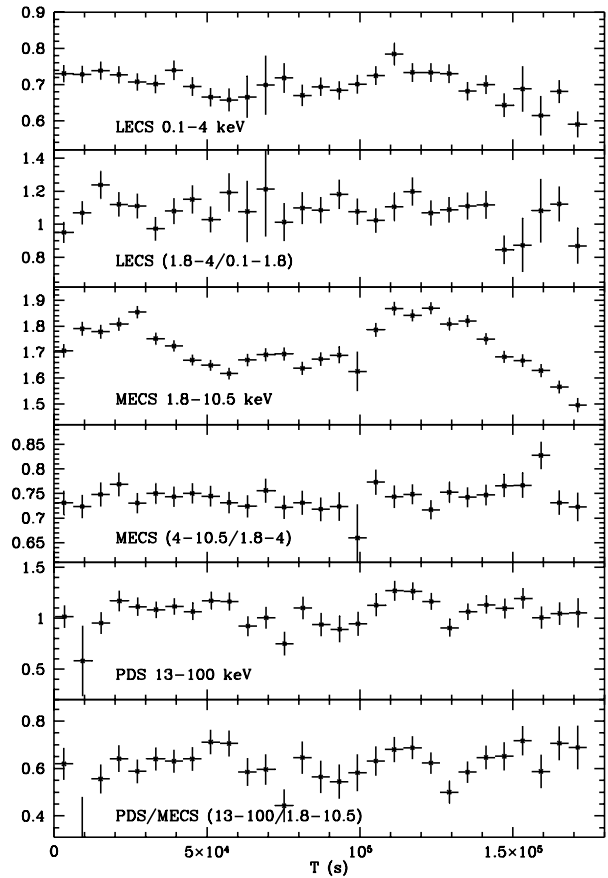}
\caption{Light and HR curves of the IC 4329A observation, in time bins of
6000 s. 
(a) Light curve in the 0.1-4.0 keV LECS band and (b) HR(1.8-4.0/0.1-1.8). 
(c) Light curve in the 1.8-10.5 keV MECS band and (d) HR(4.0-10.5/1.8-4.0).
(d) Light curve in the 13-100 keV PDS band and (f) HR(13-100/1.8-10.5).
The LECS and MECS light curves are not background subtracted.}
\end{figure}

\section{Spectral analysis}

The analysis of the integral spectral counts was performed 
using the software package XSPEC (version 10). In the first
step a ``Baseline Model Spectrum" 
(BMS) was adopted, which is composed of:
a Power Law with an exponential cut off (PL), described
by three parameters, the normalization A, the photon index
$\Gamma$ and the $e$--folding energy E$_f$ 
(A$\times$E$^{-\Gamma}\times \exp(-E/E_f)$);
a Reflection Component (RC) with two parameters, 
$r = \Omega/2 \pi$, the solid angle
fraction of a neutral, plane parallel slab illuminated by the PL photons,
and $i$, its inclination angle to the line of sight 
(module {\sc pexrav} for PL and RC together);
a uniform neutral column of gas in photoelectric absorption,
N$_H$, in addition to the adopted value of the galactic N$_{Hg}$ (module 
{\sc wabs}; in both the slab and the column the element abundances used
are the cosmic values in Anders \& Grevesse 1993);
a gaussian iron K line, with three parameters, E$_k$, $\sigma_k$
and the intensity I$_k$, also given as an EW.
Following the procedure adopted by Cappi et al. (1996),
we included in the model also two absorption edges, 
with four parameters, their energies, E$_1$ and E$_2$, 
and maximum optical depths, $\tau_1$ and $\tau_2$. 

The energy bins chosen represent about
one third of the instrumental resolution, which is a function of the energy. 
The normalization C of the PDS relative to the MECS was
adopted equal to 0.8 (Fiore et al. 1999). The ``statistical" errors
have been obtained
holding C fixed at this value, and correspond to the 90\% confidence
interval for two i. p., or
$\Delta\chi^2$=4.61. The ``systematic" errors
linked to the current $\pm$10\%  uncertainty on C 
will also be investigated.

The energies of the iron line and of the absorption edges
are reported as in the rest frame of the
host galaxy.

\subsection{The BMS fit}

The BMS fit was performed on the spectral counts from the three instruments
together. The results are given in Table 3 and illustrated 
in Fig. 2. According to the $\chi^2$ statistics, 
the BMS lies above an acceptability limit of 10\%. Moreover,
if the contribution to the $\chi^2$ from the LECS points in the interval
common to the MECS, 1.8--4 keV, is subtracted, it improves
dramatically to 115.9/118. We see no evidence in the residuals 
of a significant contribution by IC 4329 below 0.5 keV.

\begin{table}
\caption{Baseline Model Spectrum fit with cos~$i$=1}
\vspace{0.05in}
\begin{tabular}{|c|c|}
\hline
~ & ~ \cr
F(2-10~keV)$^a$ & 13.17$\pm$0.04 \cr
A$^b$ & 4.16 \cr
$\Gamma$ & 1.86$^{+0.03~(+0.03)}_{-0.03~(-0.02)}$ \cr
N$_H$~(10$^{21}$~cm$^{-2}$)$^c$ & 3.31$^{+0.26~(+0.22)}_{-0.50~(-0.08)}$ \cr
E$_f$~(keV) & 270$^{+167~(+70)}_{-80~(-40)}$ \cr
$r$ & 0.55$^{+0.15~(+0.27)}_{-0.13~(-0.20)}$ \cr
E$_k$~(keV)$^d$ & 6.48$^{+0.17}_{-0.16}$ \cr
$\sigma_k$~(keV) & 0.36$^{+0.27}_{-0.21}$ \cr
I$_k$~(10$^{-4}$~cm$^{-2}$~s$^{-1}$) & 1.57$^{+0.73~(+0.12)}_{-0.59~(-0.21)}$ \cr
EW$_k$~(eV) & 109$^{+50~(+8)}_{-41~(-14)}$ \cr
E$_1$~(keV)$^d$ & 0.73$^{+0.44}_{-0.10}$ \cr
$\tau_1$ & 0.52$^{+0.45}_{-0.37}$ \cr
E$_2$~(keV)$^d$ & 1.03$^{+0.15}_{-0.13}$ \cr
$\tau_2$ & 0.19$^{+0.16}_{-0.14}$ \cr
$\chi^2$/d.o.f. & 159.9/141 \cr
~ & ~ \cr
\hline
\end{tabular}
\begin{tabular}{c}
$^a$As observed in 10$^{-11}$ erg cm$^{-2}$ s$^{-1}$ \cr
$^b$In 10$^{-2}$ cm$^{-2}$ s$^{-1}$ keV$^{-1}$ \cr
$^c$In addition to N$_{Hg}$=4.55$\times10^{20}$ cm$^{-2}$ \cr
$^d$In the source frame \cr
\end{tabular}
\end{table}

\begin{figure}
\epsfig{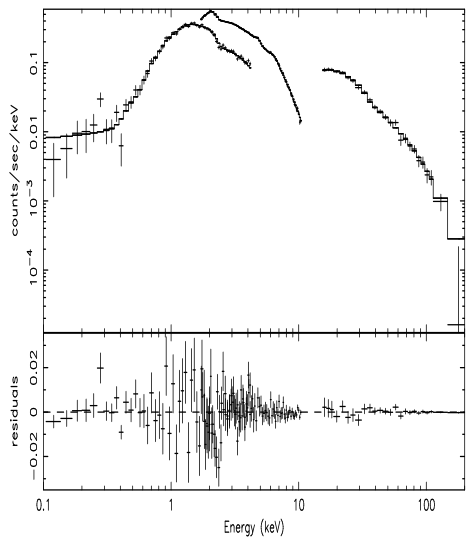}
\caption{LECS, MECS and PDS spectra with the best fit Baseline
Model Spectrum (upper panel), and residuals (lower panel).}
\end{figure}

Due to the weak dependence of the RC shape
on cos~$i$, the inclination angle is very poorly constrained, and
consequently also the value of $r$, as shown in the confidence contour plots
in Fig. 3: the best fit value of the angle is practically
0$^{\circ}$, while the 90\% upper limit read from the graph is about 
70$^{\circ}$. Lacking an objective independent
estimate of this angle (but see Sect. 4.2), we adopt as a reference 
the face--on configuration for the best fit BMS results, as given
in Table 3.

\begin{figure}
\epsfig{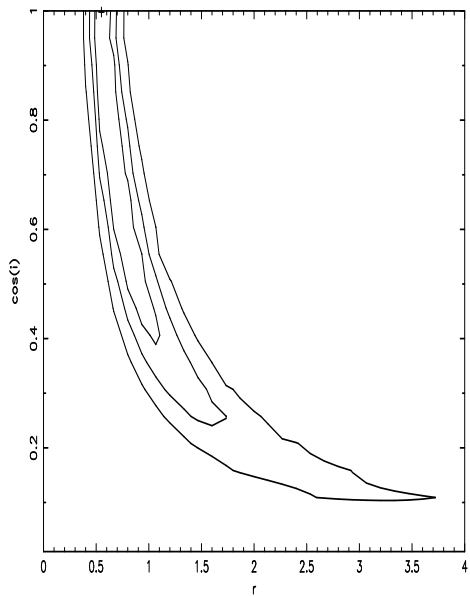}
\caption{Confidence (67, 90 and 99\%) contours of the parameters $r$ and 
cos~$i$, when the inclination angle is a free parameter in the BMS fit. }
\end{figure}

Despite the uncertainty on the angle, the evidence of a RC is remarkably
clear. However, the skewed shape of the confidence contours for the couple
of parameters $r$ and $\Gamma$ (Fig. 4) shows
that, despite the ample spectral coverage, the two parameters 
remain statistically correlated, and that the fractional 
uncertainty on $r$ is
more sensitive to the correlation than the one on $\Gamma$.

\begin{figure}
\epsfig{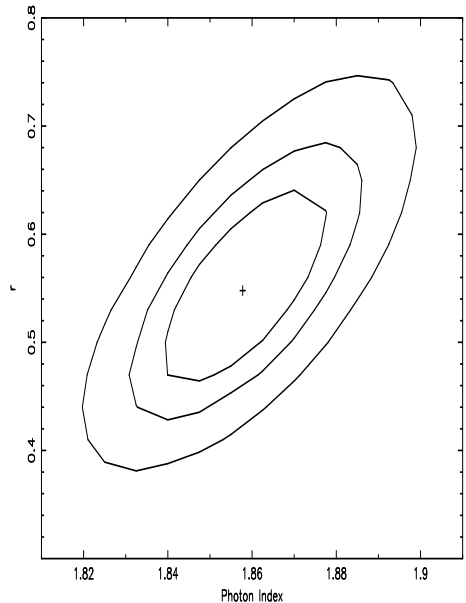}
\caption{Confidence (67, 90 and 99\%) contours of the parameters $\Gamma$ and 
$r$, in the BMS fit with cos~$i$ set equal to 1. }
\end{figure}

The next remarkable result is the evidence of the exponential
cut--off, with  rather tight constraints on the energy E$_f$. 
This is illustrated in Fig. 5, with the confidence contours
of the two parameters $\Gamma$ and E$_f$. When the BMS is fitted without
the cut--off (Fig. 6),
the residuals above 50 keV show a clear curvature,
and the $\chi^2$ is much larger,
$\Delta\chi^2$=40: according to the F--test, this difference
implies a probability less than 10$^{-3}$ that the cut--off 
is due to random fluctuations in the counts.
Notably the two broad band parameters $\Gamma$
and $r$ turn out practically the same in the two fits, thus showing
that the estimate of E$_f$ is quite well and independently constrained.

\begin{figure}
\epsfig{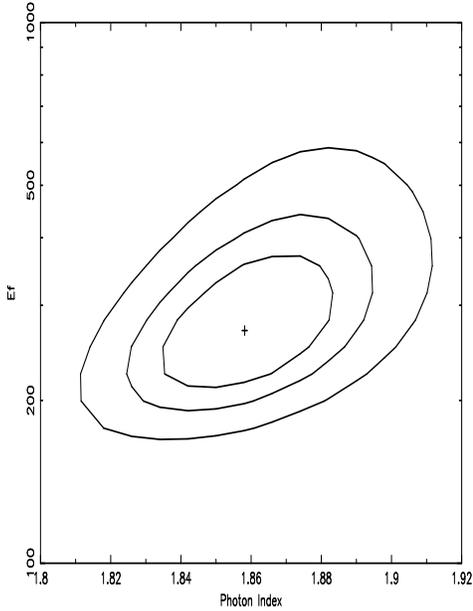}
\caption{Confidence (67, 90 and 99\%) contours of the parameters $\Gamma$ and 
E$_f$, in the BMS fit with cos~$i$ set equal to 1. }
\end{figure}

\begin{figure}
\epsfig{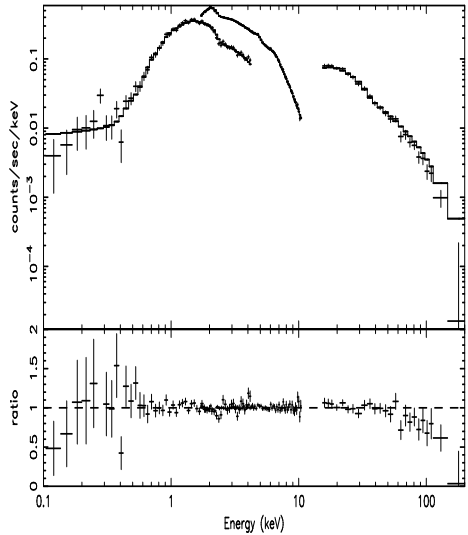}
\caption{LECS, MECS and PDS spectra with the best fit BMS without
the exponential cutoff (upper panel), and ratio of data to model
(lower panel).}
\end{figure}

The confidence contours for the energy and width of the iron K
line in Fig. 7 show that the line energy is consistent
with the neutral value and that the line appears resolved
(see Sect. 4.2).

\begin{figure}
\epsfig{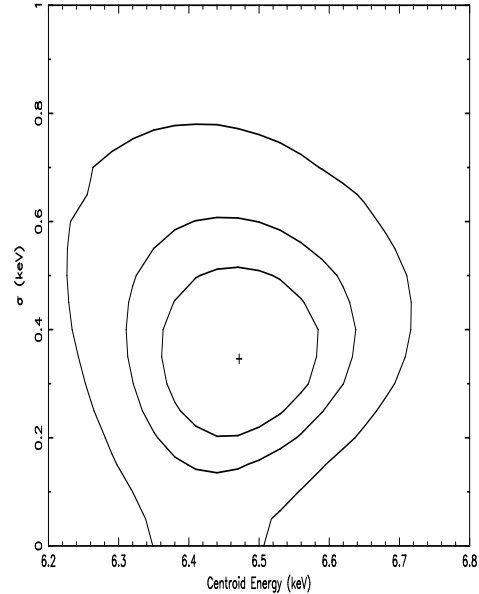}
\caption{Confidence (67, 90 and 99\%) contours of the width ($\sigma_k$) 
and energy (E$_k$) of the iron line,
in the BMS fit with cos~$i$ set equal to 1. The line energy is given in the
frame of the host galaxy. }
\end{figure}

The detection of two edges at about 0.7 and 1 keV 
confirms the presence of a warm absorber, whose physical
state will be further analyzed
in Sect. 4.3.

Finally we consider the systematic errors associated with 
the current uncertainty on the value of the PDS to MECS relative
normalization, C, mentioned above. The parameters which are significantly
affected by this uncertainty are $N_H$, $\Gamma$, E$_f$, $r$ 
and the iron line intensity
and EW. For these parameters, in Table 3, in addition to the statistical
errors from the BMS fit, we give in parentheses the systematic
error estimated from the best fit values obtained when 
C is set at 0.72 and 0.88. Notably,
while for $N_H$, E$_f$, I$_k$ and EW the systematic errors are 
comparatively small, this is not the
case for $r$, where these errors are
larger than the statistical errors, and for $\Gamma$, where they are comparable.

\subsection{Further modelling of the iron K line}

The line width is at least three times larger than in the optical
emission lines (Wandel et al. 1999), and is suggestive of 
an origin from the accretion disk. In this case it is
more natural to replace the gaussian 
with the relativistic profile, as described in the module 
{\sc diskline} in XSPEC, where a Schwarzschild black hole is adopted.
In doing that we fixed the line emission energy at 6.4 keV, the
innermost radius at 6 times R$_g$=$GM/c^2$, i.e. the innermost stable orbit,
the exponent of the radial
dependence of the emissivity at $\beta$=--2 (the use of somewhat
different values, such as $\beta$=--2.5, the average inferred by
Nandra et al. 1997, on a large sample of objects, leads to very similar 
results). The free parameters are the intensity I$_k$, the outer radius,
R$_{out}$ in units of R$_g$, of the disk region contributing 
to the line and the inclination angle;
under the assumption that most of the RC comes from the same matter,
we required the angle to be the same for the two spectral components,
hence this modification of the model leaves the number of free
parameters unchanged with respect to the BMS fit with the angle fixed at 
0$^{\circ}$ given in Table 3.

The fit with this model gives a $\chi^2$ identical to the fit
with the BMS,
and the inclination angle turns out to be 49$^{\circ~+29}_{-8}$
(notably smaller than that
of the galaxian disk).
The line parameters are R$_{out} >$158, 
I$_k$=(1.63$^{+0.71}_{-0.59})\times$10$^{-4}$ cm$^{-2}$ s$^{-1}$,
corresponding to EW$_k$=118$^{+51}_{-43}$ eV.
Compared to the BMS fit results, the values
of the other parameters in common
are practically the same, except for  the best fit value of $r$,
which, as a consequence of the inclination angle being much larger
than 0$^{\circ}$, 
turns out to be somewhat greater, $r$=0.72$^{+1.23}_{-0.18}$. 

One further fit comprises the addition of a narrow line at
6.4 keV to the previous model, with the intensity of this line
being then the only additional parameter. This addition 
leads to a marginally significant decrease in chi square,
$\Delta\chi^2$=--3.7 (93\% significance, F--test),
with the detection of a weak narrow line of EW$_{kn}$=45$^{+27}_{-31}$ eV.
The parameters of the diskline become R$_{out} <$43, EW$_k$=103$^{+56}_{-65}$ 
eV, the inclination angle $i$=45$^{\circ~+7}_{-6}$, the
reflection normalization $r$=0.72$^{+0.23}_{-0.18}$, the
others remain practically unchanged (the upper 90\% confidence
bound on $r$ is much smaller than in the previous fit because of
the substantial reduction of the corresponding bound on the angle).

If the gas with the observed N$_H$ completely surrounded the source,
the EW of the associated narrow fluorescence line would be lower
than observed by a factor about 10 
and in order to match the observed value the iron abundance should
be higher than normal by about the same factor. 
It is more plausible that the narrow line is produced in
thicker neutral matter at the outskirts or outside the accretion
disk, which should contribute also to the reflection continuum.
If we assume that the iron abundance is the same in the disk and in 
this matter, then this contribution can be inferred from the ratio of the
narrow to the total intensity, to be about 1/3.

\subsection{The warm absorber}

In order to quantify, in the frame of a single zone, equilibrium 
ionization model,
the column N$_W$ and the ionization parameter $\xi$ 
of the warm absorber, we fitted
the data with a model where the four parameters 
describing the two edges in the BMS are replaced by the two
parameters N$_W$ and $\xi$, and the {\sc XSPEC} module used is {\sc absori}. 
In this module the opacity of the
gas is based on the ionization distribution of the relevant atomic
species (same abundances as in the module {\sc wabs}), 
in a slab of Thomson--thin
gas, as a function of the ionization parameter $\xi$=$L/nR^2$ (erg cm s$^{-1}$),
where $n$ is the number density of the gas and $R$ its distance from
the ionizing source with luminosity $L$ in the interval 5 eV to 20 keV. 
We imposed the spectral slope of this source to be equal to that
of the PL. The results of the fit are given in Table 4.  
We note the improvement in $\chi^2$ with respect
to the BMS fit, $\Delta\chi^2$=--2.2 with two less parameters, 
while the parameters in common, in particular I$_k$ and $r$, 
have remained practically
unchanged. This means
that, although a single zone ionization model may often be
an oversimplification (see e.g. Reynolds 1997 and ref. therein),
in this observation, perhaps due to the limited statistics in the LECS
data as compared to ASCA data, it seems to apply rather well. 
It must be recognized, though, that $\xi$
turns out to be poorly constrained, likely because 
of the reciprocal interference of the cold and warm gas
columns, which are of comparable thickness. In Fig. 8 the best
fit model, deprived for clarity of the contribution by the cold absorbers
N$_H$ and N$_{Hg}$, illustrates the presence
of two absorption features corresponding to the two edges
detected in the BMS fit, namely  a blend of O {\sc vi} (0.67 keV) and 
O {\sc vii} (0.74 keV) edges, and a shallow trough around 1 keV, due to 
contributions by Fe L and Ne K edges.

\begin{table}
\caption{Fit with a warm absorber (cos~$i$=1)}
\vspace{0.05in}
\begin{tabular}{|c|c|}
\hline
~ & ~ \cr
$\Gamma$ & 1.88$^{+0.04}_{-0.03}$ \cr
N$_H$~(10$^{21}$~cm$^{-2}$)$^a$ & 2.62$^{+0.81}_{-1.46}$ \cr
N$_W$~(10$^{21}$~cm$^{-2}$)$^a$ & 3.05$^{+1.40}_{-1.04}$ \cr
$\xi$ & 2.4$^{+24}_{-2.0}$ \cr
E$_f$~(keV) & 313$^{+219}_{-100}$ \cr
$r$ & 0.59$^{+0.16}_{-0.13}$ \cr
E$_k$~(keV)$^b$ & 6.49$^{+0.18}_{-0.16}$ \cr
$\sigma_k$~(keV) & 0.36$^{+0.29}_{-0.21}$ \cr
I$_k$~(10$^{-4}$~cm$^{-2}$~s$^{-1}$) & 1.51$^{+0.80}_{-0.53}$ \cr
EW$_k$~(eV) & 109$^{+57}_{-31}$ \cr
$\chi^2$/d.o.f. & 157.7/143 \cr
~ & ~ \cr
\hline
\end{tabular}
\begin{tabular}{c}
$^a$In addition to N$_{Hg}$=4.55$\times10^{20}$ cm$^{-2}$ \cr
$^b$In the source frame \cr
\end{tabular}
\end{table}

\begin{figure}
\epsfig{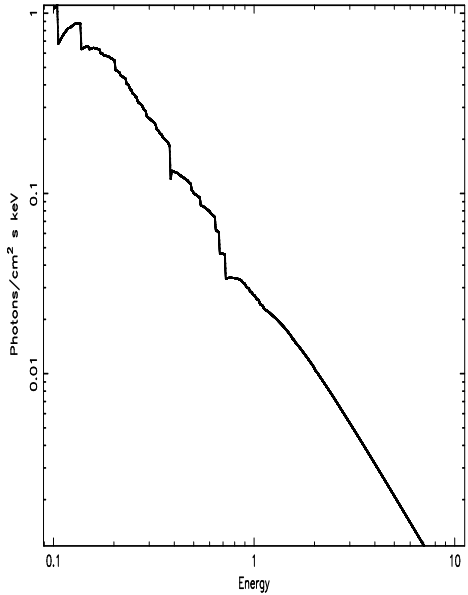}
\caption{The best fit model when the
two edges in the BMS are replaced by a column of ``warm" gas. In
this figure the cold absorber contribution is ignored for clarity.}
\end{figure}

\section{Comments and comparison with previous results}

We now comment on our results, also in comparison with
earlier results
as described in papers mentioned in Sect. 2.

If we compare the BMS fit with $i=0^{\circ}$ (Table 3) with
the {\sc diskline} 
fits ($i$=45$^{\circ}$--50$^{\circ}$),
we note that the EW of the iron line turns out to be practically the
same, about 110 eV. Adopting the ``normal" abundances in 
Anders \& Grevesse (1993), for an $\Omega = 2\pi$ geometry
the expected EW is 192 eV with $i=0^{\circ}$ (Matt et al.
1997), or about 165 eV with $i=45^{\circ}$--$50^{\circ}$ (Matt et al. 1991).
The difference in both cases with respect to the
measured value is significant and indicates that $\Omega$
is smaller than 2$\pi$. This indication is well supported by
the estimate of $r$: this is significantly smaller than one
in the BMS fit; in the {\sc diskline} fit it is only marginally
smaller than one, but the contribution to the RC by material
giving rise to the narrow line could account for about one
third of it, as noted in Sect. 4.2, thus bringing the values
of $r$ and EW from the disk in closer agreement.
An $\Omega$=2$\pi$ geometry could be retained 
by assuming either anisotropy in the primary emission pattern, or 
time delay in the response of fluorescence and 
reflection to changes in the continuum, as tentatively
found by Fiore et al. (1992).

The average flux in our observation is nearly identical to (30\% higher
than) the Ginga (ASCA) observations.
Comparison with our BMS fit of the similar one performed
by Fiore et al. (1992) on the Ginga observation shows a substantial
agreement. For a comparison with the ASCA results discussed in Cappi et
al. (1996), it is appropriate to report first the outcome of a
reanalysis of their best fit model, which is equivalent to our
BMS with $i$=0$^{\circ}$, performed with the same tools used by us
and the 90\% confidence errors for two i.p.:
$\Gamma$=1.98$^{+0.06}_{-0.09}$, $r$=2.58$^{+1.23}_{-1.11}$,
line EW=87$^{+352}_{-42}$ eV. 
We note that both Gamma and line
EW are within the errors consistent with our results, but 
that there seems to exist
an irreducible discrepancy in
the value of $r$. It is then to be remarked that, 
within the ample margins of the error on 
the iron line EW, the evidence of an internal discrepancy with respect to $r$,
noted in Sect. 2, is by no means compelling. Rather,
the comparison between the two observations indicates
that the intensity of the RC can change both in absolute and in relative
terms. The relative changes could be naturally attributed 
to a delay in the
response of the RC to variations in the intensity of the PL,
as suggested also by Cappi et al. (1996) to explain the exceptionally
large best fit value of $r$. Their suggestion is made the more
plausible by the strong, occasional outbursts of
short duration in the ASM/RXTE light curve quoted in Sect. 2.

Concerning the energy of the line, we do not confirm the ASCA
evidence in Cappi et al. (1996) that it is redshifted by
about 100 eV, their gaussian fit value being 6.30$\pm$0.07,
but with the errors in our estimate (6.48$\pm$0.17) 
we cannot exclude it either. This difference in the 
gaussian best fit energy, whatever the
cause might be, is the most likely cause of the discrepancy in inclination
angle, from the {\sc diskline} 
fit, between us and Cappi et al. (1996), who find 
an upper limit of 25$^{\circ}$ (same as found also by Mushotzky et al. 1995
and by Nandra et al. 1997 with the same observation).

In the ASCA observation Cappi et al. (1996)  found
clear evidence of two edges coinciding with  O~{\sc vii} and 
O~{\sc viii} redshifted by about 25 eV. A similar result, 
albeit different in the estimate of the
optical depths, was obtained by Reynolds (1997), who kept
the two energies fixed at their atomic values.
In the BeppoSAX observation the ionization degree of the absorber
was apparently lower, with no significant sign of the O~{\sc viii} edge.
This difference is further borne out by the estimate of the 
ionization parameter. Both Cappi et al. (1996) and 
Reynolds (1997) found 
$\xi$ equal to about 10: we find instead $\xi$ about 2.5
(although admittedly with a large uncertainty), while
the best fit values of N$_W$ are comparable. 
Since the intensity of the
source in our observation was 30\% higher than in the other,
if the strength of the ionizing flux is
the driving parameter and is proportional to the intensity of the power law, 
the reason of the difference between the two epochs
must be sought elsewhere.
In this respect it is remarkable that both Cappi et al. (1996)
and Reynolds (1997) obtain a worse $\chi^2$ with the warm absorber
model than with the two edges model.
In addition George et al. (1998),
which apply to the same observation also more sophisticated models, 
do not find a single
case which passes their acceptability limit based on the
$\chi^2$ statistics. This is suggestive
of a situation which is either more complex than the single--zone
or far from equilibrium. 
We propose the possibility that at the epoch of the ASCA observation
the ionization degree was above average and out of equilibrium,
due to a sharp transition from a high state which must have occurred
before the start of the 
observation. Non--equilibrium states are discussed in Nicastro et al. (1999a),
which typically require rather sharp transitions in the
intensity of the source. This
possibility is very attractive, in that it offers simultaneously 
an explanation for the exceptionally high value of the relative strength of
the reflection component in the ASCA data.

Lastly we stress that we have been able for the first time to 
constrain significantly the energy of the exponential cut--off
in the PL. The value we find is  close to the lower end of the ample range 
estimated by Madejski et al. (1995) by combining the
simultaneous ROSAT and OSSE (Compton GRO) observation
with the best matching (in flux) fraction of the Ginga observation.
That estimate, though, is affected by the further uncertainty
associated with  possible variations in the relative intensity of the
reflection discussed above, so is the physical modelling of the same
combination of data by Zdziarski et al. (1994).
A discussion of the physical implications of our measurement
in terms of self--consistent Comptonization models is in progress.

\section{Conclusions}

The main conclusions we draw from our analysis of this observation
of IC4329A, are the following.

a) For the first time the amplitude of the reflection component
in this galaxy has been constrained with sufficient accuracy
to draw conclusion from the simultaneously measured, and even
better constrained, EW of the iron fluorescence line. The values
of the two parameters indicate that the geometrical
factors entering the matter illumination by the primary, power
law photons imply a solid angle substantially less than 2$\pi$.
A 2$\pi$ geometry can be retained if either anisotropy in the
primary emission or geometrical lags in the response of the
reprocessed photons to changes in the primary radiation are
deemed likely to occur.

b) The power law is definitely affected by a high energy turnover,
which, if modelled as an exponential cutoff, implies an E$_f$ energy
of about 270 keV. At present, this is the fourth firm, individual determination
of a high energy turnover in Seyfert 1 spectra, after NGC 4151 
(Jourdain et al. 1992, Zdziarski et al. 1996, Piro et al. 1998), 
MCG-6-30-15 (Guainazzi et al. 1999b) and NGC 5548 (Nicastro et al. 1999b).
 
c) The state of the warm absorber in this observation appears
different from that found in the ASCA observation, in particular
the absence of a significant absorption at the O~{\sc viii} K edge
implies a lower ionization parameter, which contrasts the fact that
the intensity of the ionizing continuum, as inferred from the
direct primary emission, is actually 30\% higher.
We tentatively argue that at the epoch of the ASCA
observation the ionization state might have been 
higher than in equilibrium,
due to a transition of the continuum from a high to a low state
immediately prior to that epoch. Comfortably this same 
argument offers an explanation for the strength of the reflection
component at the same epoch, which, albeit ill constrained, seems
to have been irreducibly higher than found at the epoch of 
our observation. In other words, the comparison between the BeppoSAX
and the ASCA observations is suggestive of delay effects due
to geometrical factors in the reflection, and to relaxation
to an equilibrium state in the ionization of the absorber.


{\it Acknowledgements.} The BeppoSAX satellite is a joint Italian--Dutch program. We wish to thank the
BeppoSAX Scientific Data Center for assistance. The work was partially
supported by the Italian Space Agency, and by the Ministry for University and
Research (MURST) under grant {\sc cofin}98--02--32.

\end{document}